\documentclass[graybox]{svmult}

\usepackage{mathptmx}       
\usepackage{helvet}         
\usepackage{courier}        
\usepackage{type1cm}        
\usepackage{makeidx}         
\usepackage{graphicx}        
\usepackage{multicol}        
\usepackage[bottom]{footmisc}
\usepackage{mdwlist}
\usepackage{url}
\usepackage{subfig}
\usepackage{setspace}
\usepackage{amsmath}
\usepackage{amsfonts}
\usepackage{amssymb}

\makeindex             

\begin{document}

\title*{Using Markov Models and Statistics to Learn, Extract, Fuse,
and Detect Patterns in Raw Data}
\titlerunning{Pattern detection using Markov models}
\author{R. R. Brooks, Lu Yu, Yu Fu, Guthrie Cordone, Jon Oakley, and
  Xingsi Zhong}
\institute{Holcombe Department of Electrical and Computer Engineering,
  Clemson University, Clemson, S.C. 29643-0915 \\\email{rrb@clemson.edu}}
%
%
\maketitle

\doublespacing

\abstract*{Many systems are partially observable and stochastic in nature. We have
  derived data-driven approaches for extracting stochastic state
  machines (Markov models) directly from observed data. This chapter
  provides an overview of our approach with numerous practical
  applications. We have used this approach for inferring shipping
  patterns, exploiting computer system side-channel information, and
  detecting botnet activities. For contrast, we include a related
  data-driven statistical inferencing approach that detects and
  localizes radiation sources.}

\abstract{Many systems are partially stochastic in nature. We have
  derived data-driven approaches for extracting stochastic state
  machines (Markov models) directly from observed data. This chapter
  provides an overview of our approach with numerous practical
  applications. We have used this approach for inferring shipping
  patterns, exploiting computer system side-channel information, and
  detecting botnet activities. For contrast, we include a related
  data-driven statistical inferencing approach that detects and
  localizes radiation sources.}

\section{Introduction}
\label{sec:1}
Markov models have been widely used for detecting
patterns~\cite{sin1999network,van2007using,brooks2009behavior,
  bhanu2011side,lu2012network,lu2011botnet,fu2017botnet,fu2017stealthy}.
The premise behind a Markov models is that the current state only
depends on the previous state and that transition probabilities are
stationary.  This makes Markov models versatile, as this is a direct
result of the causal world we live in. Often, these models can only be
partially observed.  In that case, we refer to the collective
(observable and non-observable) model as a hidden Markov model (HMM).

Stochastic processes can successfully model many system signals. Some
of these systems cannot be accurately represented using a Markov model
or an HMM due to the uncertainty of input data.  One task that
benefits from stochastic signal processing is the detection and
localization of radioactive sources.  Since radioactive decay follows
a Poisson distribution~\cite{Knoll2000}, radiation measurements must
be treated as stochastic variables.  A prevalent method for localizing
radioactive sources is maximum likelihood estimation (MLE)
\cite{Gunatilaka2007,deb2013iterative,CordoneExpansion2017}, we
consider bootstrapping the MLE using estimates from a linear
regression model~\cite{CordoneMLERegression}.

In Section~\ref{sec:2}, background information about Markov models is
presented, along with references to our previous work developing
the methods presented here.  Section~\ref{sec:MM} discusses new
developments in the inference of HMMs, detection using HMMs, methods
for determining the significance of HMMs, HMM metric spaces, and
applications that use HMM-based detection.  Section~\ref{sec:SS}
analyzes radiation processes, maximum likelihood localization, and the
use of linear regression to estimate source and background
intensities.  In Section~\ref{sec:Conc}, closing remarks are
presented.

\section{Background information and previous work}
\label{sec:2}
A Markov model is a tuple $G = (S, T, P )$, where $S$ is a set of
states, $T$ is a set of directed transitions between the states, and
$P=\left\{ p(s_i,s_j)\right\} $ is a probability matrix associated
with transitions from state $s_i$ to $s_j$ such that:
\begin{equation}
\sum\nolimits_{s_j \in S} p(s_i,s_j) = 1, \quad \forall s_i \in S
\end{equation}
In a Markov model, the next state depends only on the current
state. An HMM is a Markov model with unobservable states. A standard
HMM~\cite{eddy1996hidden, rabiner1989tutorial} has two sets of random
processes: one for state transition and the other for symbol
outputs. Our model is a deterministic HMM~\cite{lu2013normalized,
  lu2012network, schwier2009pattern}, which only has one random
process for state transitions. Therefore, given the current state
and the symbol associated with the outgoing transition, the next state
is deterministic. For each deterministic HMM there is an equivalent
standard HMM and vice versa~\cite{vanluyten2008equivalence}. This
deterministic property helps us infer HMMs from observations.

Schwier et al.~\cite{schwier2011methods} developed a method for determining the optimal window size to use when inferring a Markov model from serialized data. Brooks et al.~\cite{brooks2009behavior} found that HMM confidence intervals performed better than the maximum likelihood estimate. This was illustrated through an analysis of consumer behavior. Building on this work, Lu et al.~\cite{yu2013inferring} devised a method for determining the statistical significance of a model, as well as determining the number of samples required to obtain a model with a desired level of statistical significance.

\section{Markov modeling}
\label{sec:MM}

\subsection{Infering hidden Markov models}
\label{sec:HMM}

%

HMM inference discovers the HMM structure and state transition
probabilities from a sequence of output observations. Traditionally,
the Baum-Welch algorithm~\cite{rabiner1989tutorial} is used to infer
the state transition matrix of a Markov chain and symbol output
probabilities associated with the states of the chain, given an initial
model and a sequence of symbols. As a result, this algorithm requires
the a priori structural knowledge of the Markov process that produced
the outputs.

Schwier et. al~\cite{schwier2009zero} developed the zero-knowledge HMM
inference algorithm. The assumption of HMM inference is (1) the
transition probabilities are constant; (2) the distribution of the
underlying HMM is the stationary; and (3) the Markov process is
ergodic. In essence, this means that the system is Markovian,
non-periodic and has a single strongly connected component.  The input
of the algorithm is a string of symbols, and the output is the HMM. If
the input data are not provided as a set of strings, but rather as a
set of continuous trajectories, then ~\cite{griffin2011hybrid}
explains how to change trajectories into symbolic strings.

The original algorithm only requires one parameter, $L$, as the input,
which refers to the maximum number of history symbols used to infer
the HMM state space and estimate associated probabilities. Take `abc'
as an example input string. For $L=1$, it calculates transition
probabilities $P(a|a)$, $P(a|b)$, $P(a|c)$, and so on, by
creating a state for each symbol and counting how many times the
substrings `aa', `ab', and `ac' occur in the training data and
dividing them by the number of occurences of the symbol `a'. 
The chi-squared test is used to compare the outgoing probability
distributions for all pairs of states. If two states are equivalent,
they are merged.

For $L=2$, the number of states used for representing conditional
probabilities is squared. For example, in a system that has $L=2$, it
could be the case that $P(c|a) \neq P(c|aa)$. In which case `a' and
`aa' will be different states.  The state space increases
exponentially with $L$, as does the time needed to infer the HMM. We
showed that $L$ can be determined automatically as part of HMM
inference. The idea is to keep increasing $L$ until the HMM model
stabilizes, i.e.  $HMM(L) == HMM(L+1)$. This data-driven approach
finds the system HMM with no \emph{a priori} information. Tis process
is decribed in detail in~\cite{schwier2009zero, schwier2009pattern}.

If an insufficient amount of observation data is used to generate the
HMM, the model will not be representative of the actual underlying
process. A model confidence test is used to determine if the
observation data and constructed model fully express the underlying
process with a given level of statistical
significance~\cite{yu2013inferring}. This approach calculates an lower
bound on the number of samples required. If the number of input
samples is less than the bound, more data is required. New models
should be inferred with more data and still need to be checked for confidence. This approach allows us to remove the effect of noise
in the HMM inference.

\subsection{Detection with HMM}
\label{sec:HMM_Det}

Once the HMM is inferred from symbolized data and passes the model
confidence test, it can be used to detect whether or not a data
sequence was generated by the same process. The traditional Viterbi
Algorithm~\cite{rabiner1989tutorial} finds the HMM that was most
likely to generate the data sequence by comparing probabilities
generated by the HMMs. For data streams, it is unclear what sample
size to use with the Viterbi algorithm. Also as data volume increases,
the probability produced by the Viterbi algorithm decreases
exponentially, and may suffer floating point
underflow~\cite{brooks2009behavior}. To remedy this, confidence
intervals (CIs) are used~\cite{brooks2009behavior}. With this
approach, the certainty of detection increases with the number of
samples and the floating point underflow issues of the Viterbi
Algorithm is eliminated~\cite{brooks2009behavior}. The CI approach can
determine, for example, whether or not observed network traffic was
generated by a botnet HMM.

Given a sequence of symbolized traffic data and a HMM, the CI method
in~\cite{brooks2009behavior} traces the data through the HMM and
estimates the transition probabilities and confidence intervals. This
process maps the observation data into the HMM structure. It then
determines the proportion of original transition probabilities that
fall into their respective estimated CIs. If this percentage is
greater than a threshold value, it accepts that the traffic data
adequately matches the HMM.

Lu et al.~\cite{lu2012network} used an alternative approach. Instead
of estimating transition probabilities and corresponding CIs, they
calculated the state probabilities, which are the proportion of time
the system stays in a specific state, and their corresponding
CIs. Once all state probabilities and CIs were estimated for the
observation sequence, they determined whether each state is
within a certain confidence interval. For the whole HMM, the
proportion of states was obtained whose estimated state probability
matched the corresponding confidence intervals. If an observation
sequence was generated from the HMM, it would follow the state
transitions of the underlying stochastic process that the HMM
represents, and its state probabilities would converge to the
asymptotic state probabilities if the sequence length was large
enough. Therefore, more states would match their estimated
CIs. Generally, a sequence that is generated by the HMM would have a
high proportion of matching probabilities, while a sequence that is
not an occurrence of the HMM will have a low proportion of matching
probabilities. Similar to the detection approach
in~\cite{brooks2009behavior}, a threshold value can be set for this
proportion of matching states, to determine whether the observation
sequence matches with the HMM.

In~\cite{brooks2009behavior}, receiver operating characteristic (ROC)
curves were used to find optimal threshold values. By varying the
threshold from 0\% to 100\% (0\% threshold means we accept everything
and we will have a high false positive rate. 100\% threshold means we
reject everything and we will have a low true positive rate), they
progressively increased the criteria for acceptance. Using the ROC
curve drawn from detection statistics with different thresholds, the
closest point to (0,1) was found, which represents 0\% false positive
rate and 100\% true positive rate. This considered the trade-off
between true positive and false positive rates. Therefore, the
corresponding threshold of that point was the optimal threshold value.

\subsection{Statistically significant HMMs}
\label{sec:HMM_Sig}
\begin{figure}[h]
	\centering
	\includegraphics[scale=.5]{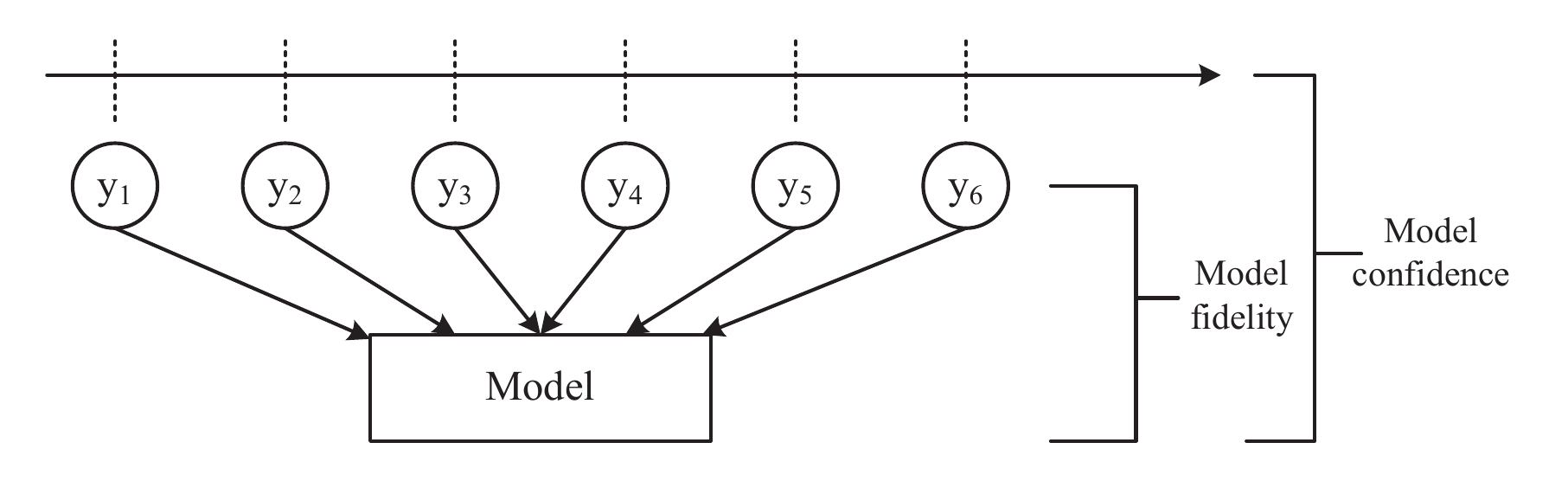}
	\caption{Hierarchy of the process, observations, and model showing the relationship between
model fidelity and model confidence (adopted from~\cite{yu2013inferring}).}
	\label{fig:fidelity}
\end{figure}

In~\cite{yu2013inferring}, we address the problem of how to ensure the inferred HMM accurately represents the underlying process
that generates the observation data.  The measure of the accuracy is
known as \textit{model confidence}, which means the degree to which a
model represents the underlying process that generated the training
data.  As illustrated in Figure~\ref{fig:fidelity}, model confidence
is different from \textit{model fidelity}, where the latter refers to
the agreement between the inferred model and the training data.

Mathematically, we calculate an upper bound on the number of samples
required to guarantee that the constructed model has a given level
significance.  In other words, we have shown how to determine within a
given level of statistical confidence ($\alpha$) if a ``known
unknown'' transition does not occur, given two user-defined thresholds
$\epsilon$ and $\alpha$.  The parameter $\epsilon$ determines the
minimum probability that transitions with probabilities no less than
$\epsilon$ should be included in the constructed model.  The parameter
$\alpha$ is the confidence level that shows the accuracy of the model
result.

For each state, $s$, we use the $z$-test~\cite{bowerman1990linear} to
determine if the inferred model includes all transitions with
probabilities no smaller than
$\gamma_{s}^{\epsilon} = {\epsilon}/{\pi_s}$ with desired significance
$\alpha$, where $\pi_s$ is the asymptotic probability of $s$ as the
constructed Markov model is irreducible.  Concretely, we test the null
hypothesis $H_0$: $\overline{X_{s}^U} = \gamma_{s}^{\epsilon}$ against
the alternative hypothesis $H_1$:
$\overline{X_{s}^U} < \gamma_{s}^{\epsilon}$, where
$\overline{X_{s}^U}$ denote the sample average of the probability of
an unobserved transaction of state $s$.  We use
$\gamma_{s}^{\epsilon} - \overline{X_{s}^U}$ as a test statistic,
rejecting the null hypothesis if
$\gamma_{s}^{\epsilon} - \overline{X_{s}^U}$ is too large.  We have
not observed the transition, thereby $\overline{X_s^U} = 0$.  We
fail to reject $H_0$ if we need to collect more data.  Otherwise, we say that sufficient data has been collected.  If we fail to reject the null
hypothesis for all states, the $z$-test also finds the minimum amount
of training samples needed for transaction transitions with
probabilities no smaller than $\gamma_{s}^{\epsilon}$ to be detected
with desired level of significance $\alpha$.

To use the $z$-test in this manner, we propose a simple algorithm to
perform on-line testing of the observation sequence. The algorithm
determines if a constructed model statistically represents a data
stream in the process of being collected.  We first collect a sequence
of observation data $\mathbf{y}$ of some length $D$ and construct a
model from the collected data. With the constructed model, we
determine the $z$-statistics and find if the experimental statistic
provides $100 \cdot (1-\alpha)\%$ confidence that a transition with
probability $\epsilon$ does not occur. If $\mathbf{y}$ is not
sufficiently long, we will be unable to construct a model from the
data; additional data should be gathered.  The algorithm is provided
in~\cite{yu2013inferring}.

\subsection{HMM metric space}
\label{sec:HMM_Met}
A metric is a mathematical construct that describes the similarity (or
difference) between two models.  It is useful to know if two processes
are the same, except for rare events (e.g., events occurring once in a
century), since we would typically consider them functionally
equivalent.  Eliminating duplicate models can reduce system complexity
by decreasing the number of models used to analyze using observation data.
Grouping similar models can increase the number of samples available
for model inference, leading to higher fidelity system
representations.
\begin{figure}[h]
	\centering
	\includegraphics[scale=.7]{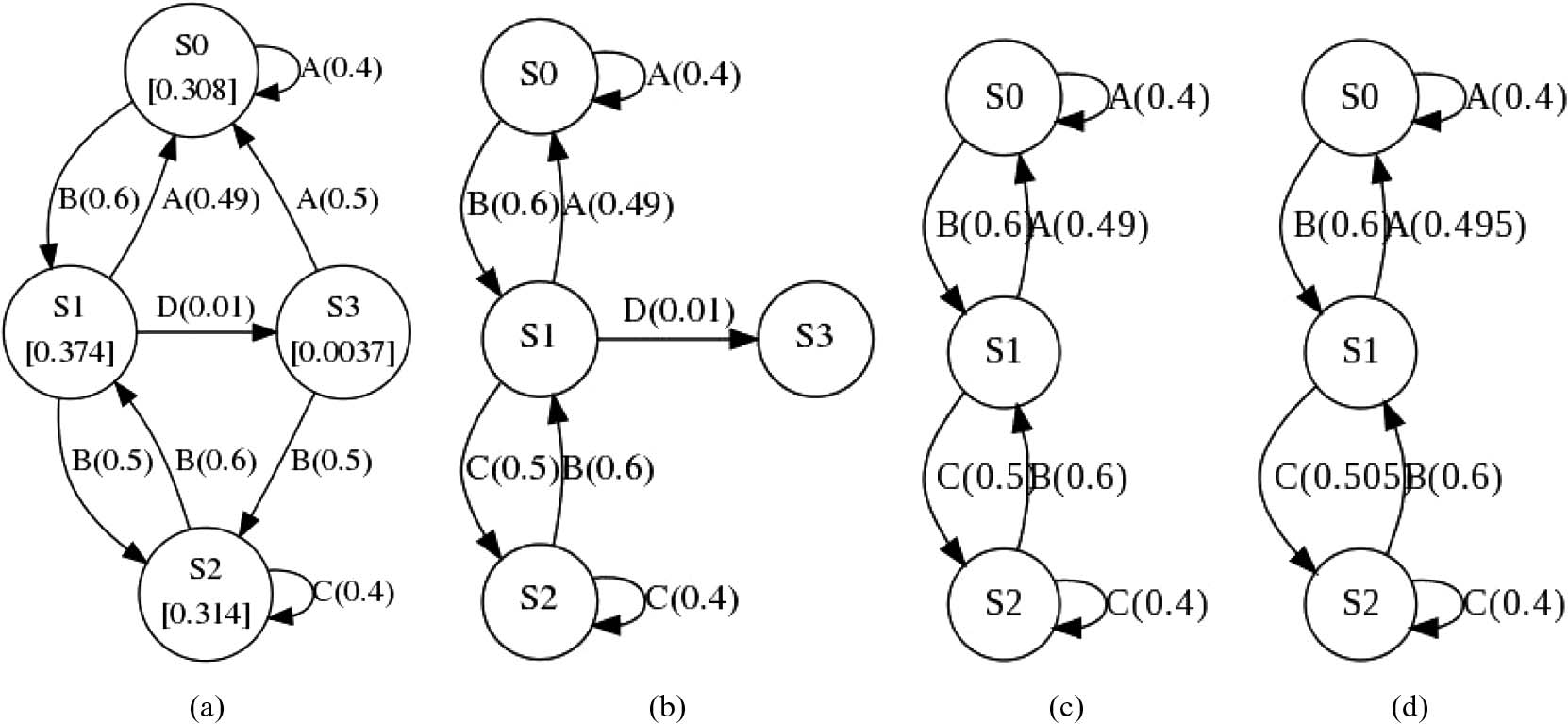}
	\caption{Process to remove transitions: (a) original model (asymptotic state probabilities are in square brackets); (b) initial step with $P_{th} = 0.002$; (c) removal of absorbing states; and (d) resulting model with rescaled probabilities. (adopted from~\cite{lu2013normalized}).}
	\label{fig:metric}
\end{figure}
To determine if two deterministic HMMs $G_1$ and $G_2$ are equivalent,
we first let $G_1$ generate an observation sequence $\mathbf{y}$ of
length $D$ and then run $\mathbf{y}$ through $G_2$.  Using frequency
counting, we obtain a set of sample transition probabilities for each
transition of $G_2$, denoted by $T'_2$.  The probabilities of
all outgoing transitions conditioning on each state follow multinomial
distribution, which are then approximated by a set of binomial
distributions in this context.  This allows us to use the standard
$\chi^2$-test to compare if the two sets of transition distributions
$\{T'_2, T_2\}$ are equivalent with significance
$\alpha$~\cite{kutner2004applied}, where $T_2$ is the set of
transitions probabilities of $G_2$.  We define equivalence
$(G_1\equiv G_2)$ as $G_1$ and $G_2$ accepting the same symbol
sequences with a user-defined statistical significance $\alpha$.  The
corresponding algorithm is provided in~\cite{lu2013normalized}.  Note
that, if there does not exist a path across $G_2$ that yields
observation sequence $\mathbf{y}$, we can immediately conclude that
$G_1$ and $G_2$ are not equivalent.

If $G_1$ and $G_2$ are not equivalent, we quantify the distance
between them as how much their statistics differ with significance
$\alpha$.  Given a probability $0\leq P_{th}\leq 1$, we prune all
transactions with probability no greater than $P_{th}$ from both
models and then re-normalize the probabilities of remaining
transitions for each state to obtain two sub-models.  The pruning
process is illustrated in Figure~\ref{fig:metric} using a simple model
with threshold value $P_{th} = 0.002$. The distance between the two
models is defined as the minimum $P_{th}$ that yields two equivalent
sub-models.  To determine this distance, we progressively remove the
least likely events from both models until the remaining sub-models
are considered equivalent.  Specifically, starting from $P_{th}=0$, we
prune all transactions with probability no greater than $P_{th}$, and
then re-normalize the probabilities of remaining transactions for each
state.  We gradually increase $P_{th}$ and repeat the pruning
operation for each $P_{th}$ until the remaining sub-models are
considered equivalent with the predefined confidence level $\alpha$.
Let $d(G_1,G_2)$ denote the statistical distance between $G_1$ and
$G_2$, then $d(G_1,G_2)$ is equal to the stopping value of $P_{th}$.  We
show in~\cite{lu2013normalized} that $d(G_1,G_2)$ is a metric as it
fulfills all the necessary conditions as a metric.

Since the $\chi^2$-test is used, we need to ensure that enough samples
($D$) are used for transition probabilities to adequately approximate
normal distributions.  In order to approximate the binomial
distribution as normal distribution, the central limit theorem is used
to calculate the required sample size.

This is similar in spirit to Kullback--Leibler divergence (KLD);
however, unlike KLD, our approach is a true metric.  It does not have
KLD's limitations and provides a statistical confidence value for the
distance. We illustrated the performance of the metric by calculating
the distance between different, similar, and equivalent models.
Compared with KLD measurement, our approach is more practical and
provides a true metric space.

\subsection{HMM detection applications}
\label{sec:HMM_DA}
We have used this approach to:
\begin{itemize*}
\item Track shipping patterns in the North Atlantic;
\item Identify protocol use within encrypted VPN traffic;
\item Identify the language being typed in an SSH session;
\item Identify botnet traffic;
\item Identify smart grid traffic within encrypted tunnels and disrupt
  data transmission;
\item Create transducers that transform the syntax of one network
  protocol into another one; and
\item De-anonymize bitcoin currency transfers.
\end{itemize*}
\subsubsection{Track shipping patterns in the North Atlantic}

Symbolic Transfer Functions (STF) were developed for modeling
stochastic input/output systems whose inputs and outputs are both
purely symbolic. Griffin et al.~\cite{griffin2011hybrid} applied STFs
to track shipping patterns in the North Atlantic by assuming the
input symbols represent regions of space through which a track is
passing, while the output represents specific linear functions that
more precisely model the behavior of the track. A target's behavior is
modeled at two levels of precision: the symbolic model provides a
probability distribution on the next region of space and behavior
(linear function) that a vehicle will execute, while the continuous
model predicts the position of the vehicle using classical statistical
methods. They collected over 8 months worth of data for 13 distinct
ships, representative of a variety of vessel classes (including cruise
ships, Great Lakes trading vessels, and private craft). The STF algorithm
was used to produce a collection of vessel track models. The models were tested for their prediction power over the course of three
days.

\subsubsection{Identify protocol use within encrypted VPN traffic}\label{sec:vpn}

\begin{figure}[h]
	\centering
	\includegraphics[width=3.5in]{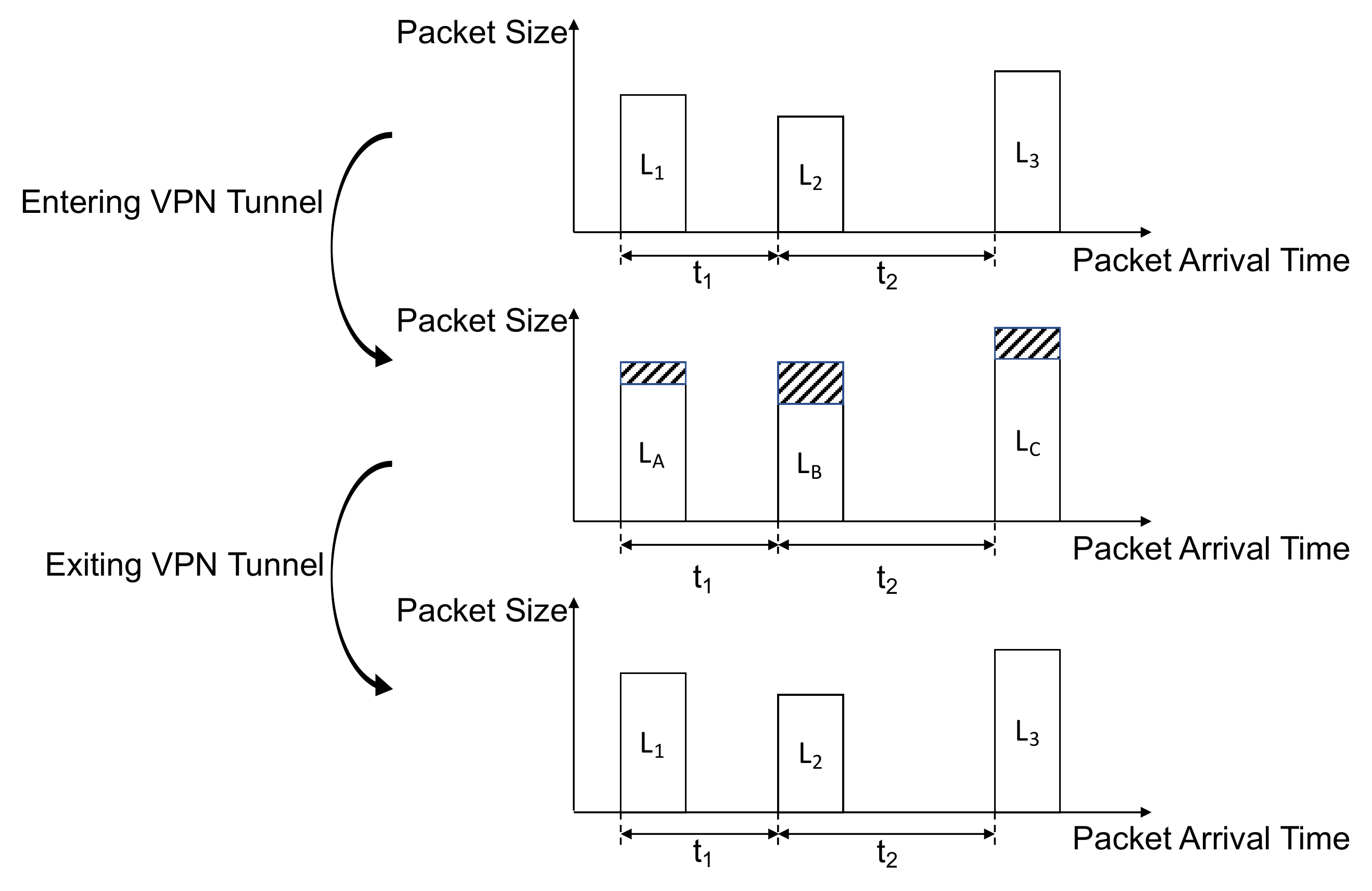}
	\caption{Timing and size side-channels in a VPN tunnel.}
	\label{fig:sidechannels}
\end{figure}

A VPN is an encrypted connection established between two private
networks through the public network. Packets transferred through a VPN
tunnel have the source and destination IP of the private networks,
which are not always the final destination of the packet. The
destination IPs for each packet are encrypted and cannot be seen by
any third party. A typical VPN implementation encrypts the packets
with little overhead and pads all packets in a given size range to the
same size. Thus, the timing side-channel and packet size side-channel
can be used to identify the underlying protocol even after
encryption. These side-channels are shown in
Figure~\ref{fig:sidechannels}. In~\cite{zhong2015twoPMU}, a
synchrophasor network protocol is identified in an encrypted VPN
tunnel.

\subsubsection{Identify the language being typed in an SSH session}

SSH encryption is mathematically difficult to break, but the implementation is vulnerable to side-channel analysis. In an interactive SSH session, users' keystroke-timing data are associated with inter-packet delays.
In~\cite{harakrishnan2011side}, the inter-packet delays are used to determine the language used by the victim. This is achieved  by comparing the observed data with the HMM of known languages. In~\cite{Song:2001:TAK:1251327.1251352}, the timing side-channel is used to extract the system password from interactive SSH sessions. 

\subsubsection{Identify botnet traffic}
Botnets are becoming a major source of spam, distributed denial-of-service attacks (DDoS), and other cybercrime. 
Chen et al~\cite{lu2011botnet} used traffic timing information to detect the centralized Zeus botnet. The reasons to
use timing information are: (1) inter-packet timings relate to the command and control processing time of the
botnet; (2) the bots periodically communicate with the command and control (C\&C) server for new commands or new data; (3) the inter-packet delays filter out constant communication latencies; and (4) it does not require reverse engineering the malware binaries or encrypted traffic data. An HMM is inferred from inter-packet delays of Zeus botnet C\&C communication traffic. Using the behavior detection method with confidence intervals (CIs) of HMMs~\cite{brooks2009behavior}, they were able to detect whether or not a sequence of traffic data is botnet traffic. Experimental results showed this approach can properly distinguish wild botnet traffic from normal traffic. 

\subsubsection{Identify smart grid traffic within encrypted tunnels and disrupt data transmission}

 In a power grid, Phasor Measurement Units (PMU) send measurement data (or synchrophasor data) over the Internet to a control center for closed loop control. Any unexpected change in the variance of the packet delay can cause instability in the power grid and even blackouts. PMUs are deployed at critical locations and are usually protected by security gateways. 

\begin{figure}[t]
\centering
\subfloat[]{\includegraphics[width=3.2in]{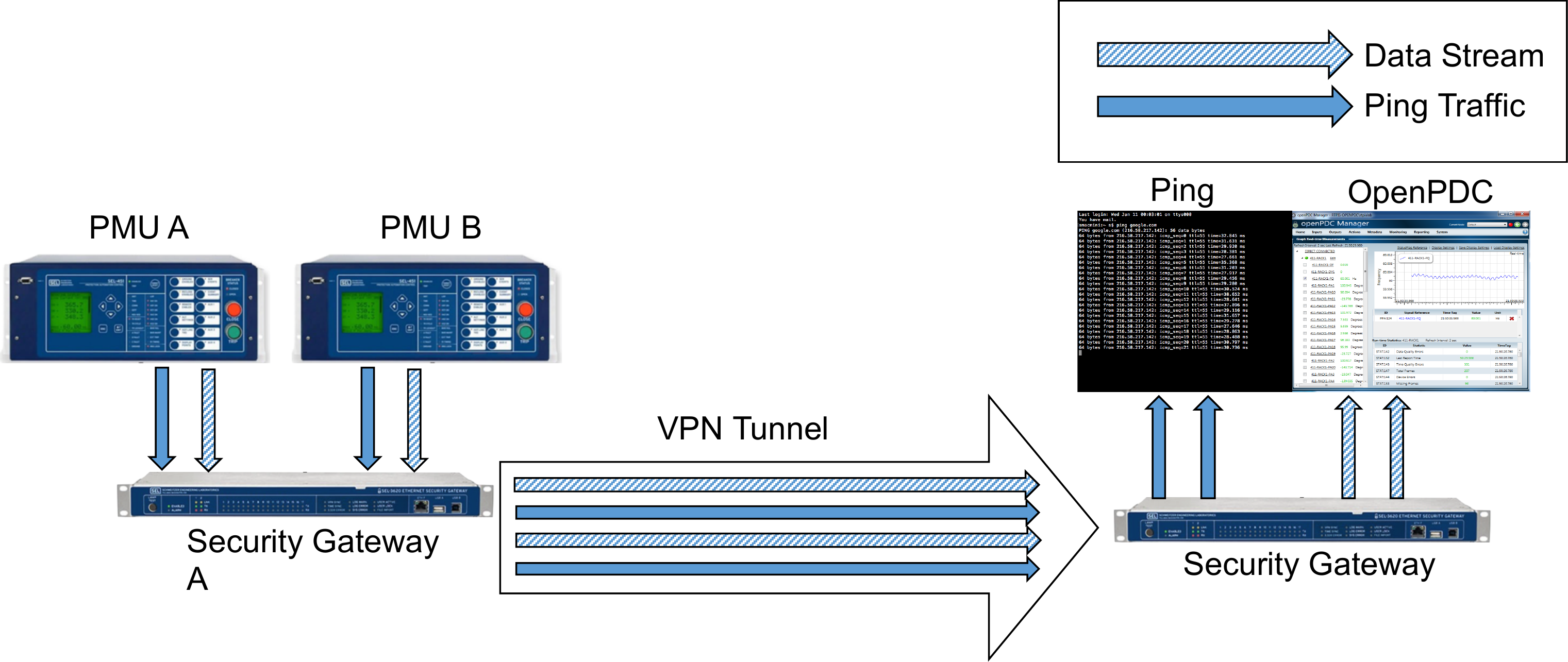}
\label{fig:attack0}}
\hfil
\subfloat[]{\includegraphics[width=3.2in]{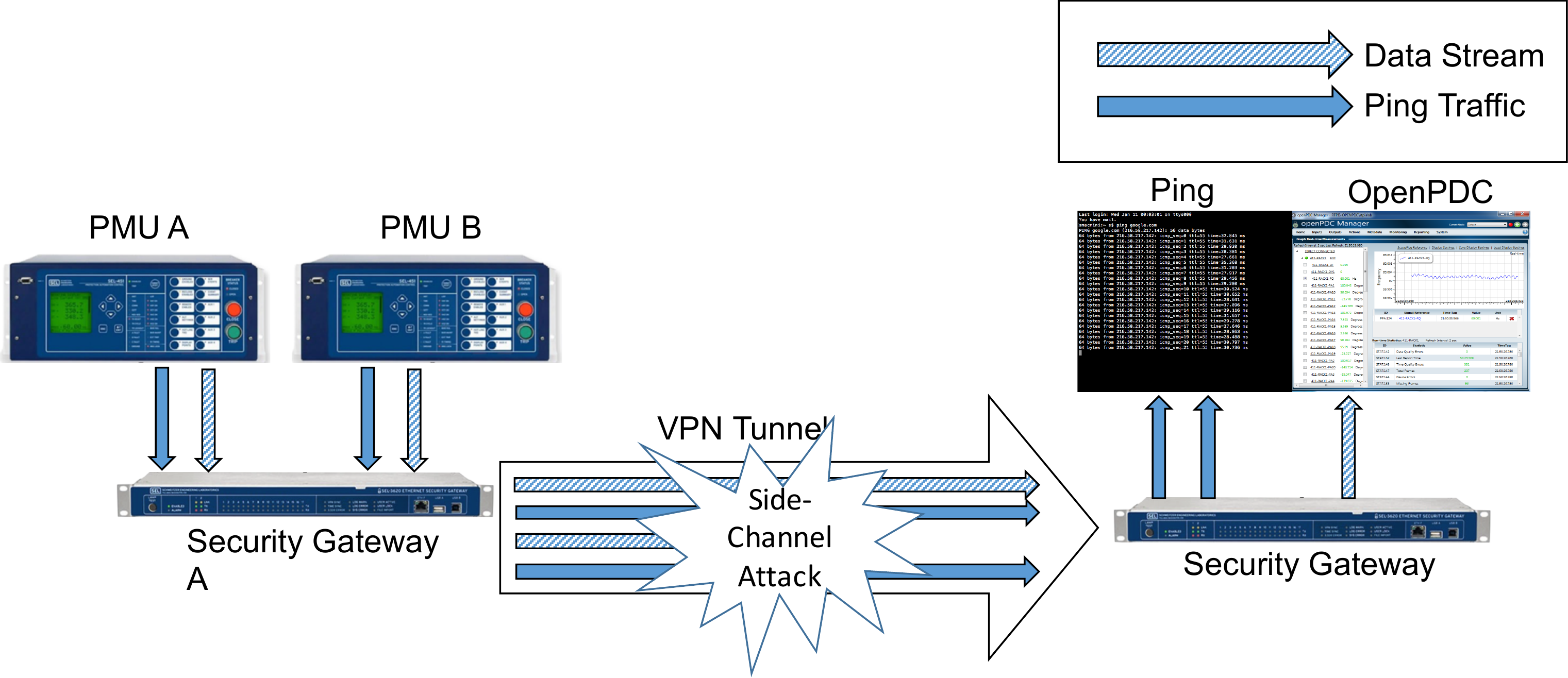}
\label{fig:attack1}}
\caption{An example of a DoS attack in a VPN tunnel: (a) A VPN network carries Ping traffic and data streams from two PMUs to control center; (b) During a DoS attack, a PMU data stream within the VPN tunnel is identified and dropped.}
\label{fig:attackexample}
\end{figure}

 The use of security gateways and VPN tunnels to encrypt PMU traffic eliminates many possible vulnerabilities~\cite{beasley2014survey,zhong2015cyber}, but these devices are still vulnerable to Denial of Service (DoS) attacks that exploit a side-channel vulnerability. In~\cite{zhong2017ETCI}, encrypted PMU measurement traffic is identified and dropped through an exploitation of side-channel vulnerabilities. The underlying protocol is detected as described in Section~\ref{sec:vpn}, and the target protocol's packets are dropped. The attack leads to unstable power grid operations. Figure \ref{fig:attackexample} illustrates such an attack. This attack is tough to detect because only the measurement traffic is dropped, and all other traffic is untouched (e.g. ping or SSH traffic). From the victim's point of view, all systems works fine, but no measurement traffic is received.

\subsubsection{Botnet Domain Generation Algorithm (DGA) modeling and detection}
Botnets have been problematic for over a decade. 
In response to take-down campaigns, botmasters have developed techniques to evade detection. One widely used evasion technique is DNS domain fluxing. Domain names generated with Domain Generation Algorithms (DGAs) are used as the \textit{rendezvous points} between botmasters and bots. In this way, botmasters hide the real location of the C\&C servers by changing the mapping between IP addresses and domain names frequently. Fu et al.~\cite{fu2017stealthy} developed a new DGA using HMMs, which can evade current DGA detection methods (Kullback-Leibler distance, Edit distance, and Jaccard Index)\cite{yadav2010detecting} and systems (Botdigger~\cite{zhang2016botdigger} and Pleiades~\cite{antonakakis2012throw}). The idea is to infer an HMM from the entire space of IPv4 domain names. Domain names generated by the HMM are guaranteed to have the same lexical features as the legitimate domain names. With the opposite idea, two HMM-based DGA detection methods were proposed~\cite{fu2017botnet}. Since the HMM expresses the statistical features of the legitimate domain names, the corresponding Viterbi path of a given domain name can be found, which indicates the likelihood that the domain name is generated by the HMM. The probability returned by the HMM is a measure of how consistent the domain name is with the set of legitimate domain names.

\subsubsection{A covert data transport protocol}
Protocol obfuscation is widely used for evading censorship and surveillance and hiding criminal activity. Most firewalls use DPI to analyze network packets and filter out sensitive information. However, if the source protocol is obfuscated or transformed into a different protocol, detection techniques won't work~\cite{zhong2015stealthy}. 
Fu et al.~\cite{fu2016covert} developed a covert data transport protocol that transforms arbitrary network traffic into legitimate DNS traffic in a server-client communication model. The server encodes the message into a list of domain names and register them to a randomly chosen IP address. The idea of the encoding is to find a unique path in the HMM inferred from legitimate domain names, which is associated with the message. The client does a reverse-DNS lookup on the IP address and decodes the domain names to retrieve the message. Compared to DNS tunneling, this method doesn’t use uncommon record types (TXT records) or carry suspiciously large volume of traffic as DNS payloads. On the contrary, the resulting traffic will be normal DNS lookup/reverse-lookup traffic, which will not attract attention. The data transmission is not vulnerable to DPI.

\subsubsection{Bitcoin Transaction Analysis}
Bitcoin has been the most successful digital currency to date.  One of the main factors contributing to Bitcoin's success is the role it has found in criminal activity. According to Christin~\cite{christin2013traveling}, in 2012 the Silk Road (a popular dark web marketplace) was handling 1.2 million USD worth of Bitcoin transactions each month.  This is largely attributed to Bitcoin's appeal--pseudonymity, which disassociates users with the publicly available transactions.

\begin{figure}[h]
	\centering
	\includegraphics[scale=.4]{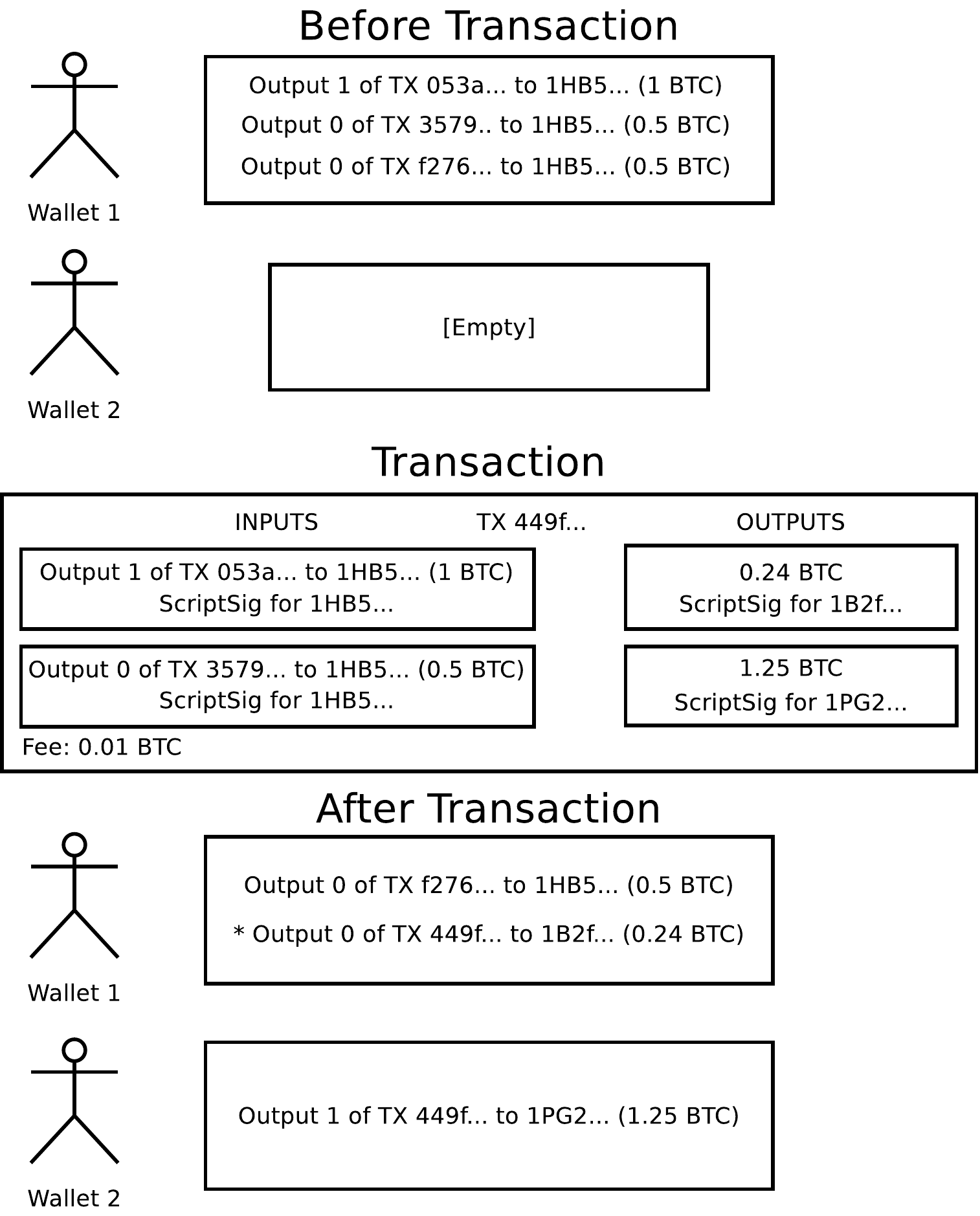}
	\caption{An example of a Bitcoin transaction found in the public blockchain.}
	\label{fig:blockchain}
\end{figure}

It is natural to consider financial transactions as a Markovian process since each transaction is governed solely by previous states. All Bitcoin transactions can be publicly viewed on the blockchain, as shown in Figure~\ref{fig:blockchain}. This motivates the notion that transactions can be represented by a Markov model. However, since Bitcoin is pseudonymous, there are hidden states that correspond to the Bitcoin users.  The process of grouping the observable transactions with their respective users infers the underlying HMM.

Theoretically, this HMM would render traditional Bitcoin money laundering useless since existing Bitcoin laundering techniques are reminiscent of a shell game\footnote{\url{https://en.wikipedia.org/wiki/Shell_game}}. Current research is focused on inferring the HMM from the Bitcoin blockchain, leveraging existing research that identifies transient transactions used for change functionality~\cite{barber2012bitter,ober2013structure}. 

\section{Stochastic Signal Processing for Radiation Detection and Localization}
\label{sec:SS}

\subsection{Radiation Processes}
\label{sec:RP}
The detection and localization of radioactive sources, especially those that emit ionizing gamma radiation, is a problem that is of great interest for national security~\cite{uscongress2010}.  Such a task is not simple due to the physics of radiation signal propagation, the stochastic nature of radiation measurements, and interference from background noise.  Assuming a uniform propagation medium between the source and location of measurement, as well as a negligible attenuation due to the medium, a simple radiation propagation model is given by

\begin{equation}
I=\frac{A}{x^2}+B ,
\label{propagationmodel}
\end{equation}

\noindent
where $I$ is the total radiation intensity at the measurement location, $A$ is the intensity of the radioactive source, $x$ is the distance from the source to the measurement location, and $B$ is the intensity of the background radiation at the measurement location.

\begin{figure*}[b]
	\centering
	\includegraphics[width=0.45\linewidth]{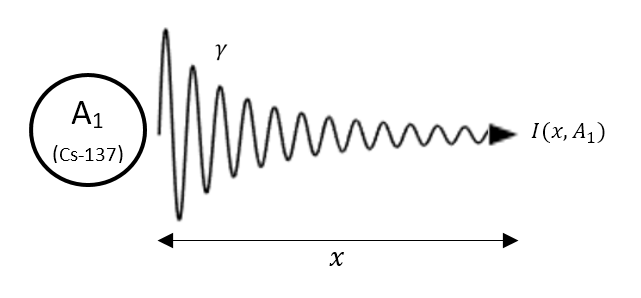}
	\caption{The intensity of a radiation signal decreases exponentially with distance when moving through a uniform medium}
	\label{fig:radiation_propagation}
\end{figure*}

It is evident from (\ref{propagationmodel}) that propagation of the signal from a radioactive point source is governed by the inverse-square law, which states that the intensity of the signal is inversely proportional to the distance from the source.  Figure \ref{fig:radiation_propagation} provides a visual example of the effect the inverse-square law has on the intensity of a radiation signal.  

\begin{figure}[h]
	\centering
	\includegraphics[width=\linewidth]{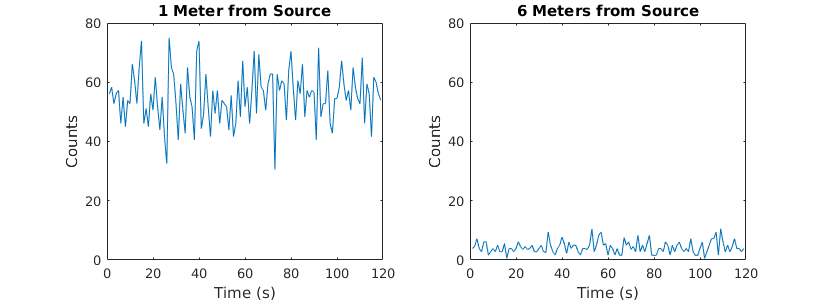}
	\caption{Real measurements of a 35 $\mu Ci$ Cs-137 source at various distances (IRSS datasets~\cite{irss})}
	\label{fig:real_counts_comparison}
\end{figure}

Ionizing radiation is commonly measured using scintillation counters, which integrate the number of times that an incident gamma particle illuminates a scintillation material over a given time period.  The number of incidents integrated over each time period are referred to as `counts.'  Due to the physics of radioactive decay, the measurements of scintillation counters are Poisson random variables~\cite{Knoll2000}.  As a result, single radiation measurements are unreliable since the measurement of a high intensity source will have a proportionally high variance.  An example of the difference in variance between a high intensity and low intensity signal is given in Figure \ref{fig:real_counts_comparison}.  Observe that the high intensity signal recorded one meter from the source has a much higher range of values than the low intensity signal recorded six meters from the source.  

\subsection{Maximum Likelihood Estimation}
\label{sec:MLE}
Basic radiation detection methods for national security involve the deployment of one or two sensors in the form of portal monitors located at choke points along the road.  While simple and reliable for detection, the use of portal monitors is not practical for several locations, such as a widespread urban environment with complex road networks~\cite{Brennan2004}.  Recently, the major focus of research has been on the use of distributed networks of detectors, which are much more suitable for the detection and localization of radiation sources over a widespread area.  With a distributed detector network, the detection and localization of radiation sources becomes a complex problem requiring the fusion of large amounts of stochastic data.  One of the most common fusion methods used for this purpose is Maximum Likelihood Estimation~\cite{Gunatilaka2007,chin2008,vilim2009radtrac}.

\begin{figure}[t]
	\centering
	\subfloat[]{\includegraphics[width=0.45\linewidth]{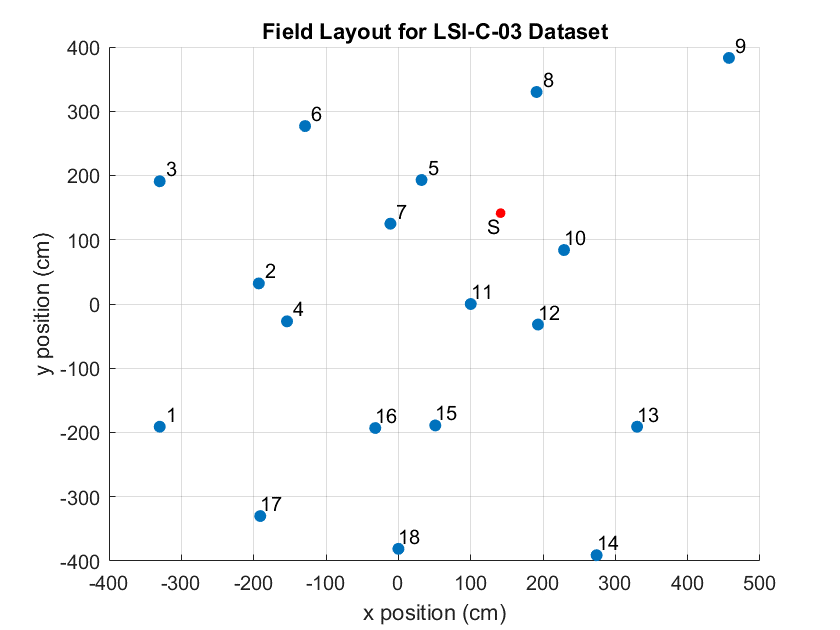}
		\label{fig:c03_layout}}
	\hfil
	\subfloat[]{\includegraphics[width=0.45\linewidth]{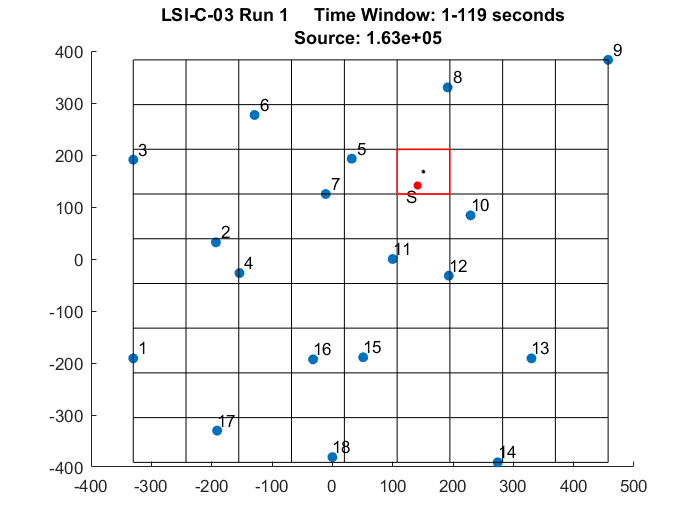}
		\label{fig:c03_mle}}
	\caption{An example of a low resolution MLE localization: (a) Layout of a detector field from IRSS datasets \cite{irss}.  Detectors are blue dots indexed by numbers and the radiation source is a red dot tagged with an 'S'. (b) A coarse MLE localization has been performed over the detector field.  The center of each grid region was used as input parameters for (\ref{maximumlikelihood}).  The grid region with a red border and a black dot at the center is the area selected by the search.}
	\label{fig:mle_example}
\end{figure}

In short, Maximum Likelihood Estimation (MLE) is a search over all possible parameters of the target radiation source.  While the number of parameters depends on the radiation model, they commonly include the horizontal and vertical coordinates of the source and the radioactive intensity of the source.  Each combination of possible parameters are plugged into a likelihood function, which gives the probability that a source with the given parameters cause each detector in the field to have their current measurements.  The parameters that generate the highest likelihood are selected as the maximum likelihood estimate of the source. An example likelihood function based on the radiation model in (\ref{propagationmodel}) is given by

\begin{equation}
L(\theta)=\sum_{i=1}^{N_d}\left(\ln{I_i}\sum_{j=1}^{w}\left(c_{ij}-1\right)\right) ,
\label{loglikelihood}
\end{equation}

\noindent
where $N_d$ is the number of detectors being used for the localization, $w$ is the length of the time window, $c_{ij}$ is the measurement for the $i$th detector at timestep $j$, and $\theta$ is the vector of input source parameters, which are used to generate the the intensity at the $i$th detector, $I_i$, using (\ref{propagationmodel}).  The maximization of the likelihood function is then given as

\begin{equation}
\hat{\theta}_{ml} = \arg{\max{L(\theta)}} ,
\label{maximumlikelihood}
\end{equation}

\noindent
where $\hat{\theta}_{ml}$ is the maximum likelihood estimate of the source parameter vector.  A visual example of an MLE localization is shown in Figure \ref{fig:mle_example}.  Note that the likelihood function in (\ref{loglikelihood}) is a simplification of the logarithm of the joint Poisson probability for the likelihood of the measurements at each individual detector~\cite{CordoneExpansion2017}.

The primary criticism of MLE localization is that it is computationally intensive, requiring a brute force search over all parameters.  A common method for reducing the total computational load of the MLE search is the use of an iterative search~\cite{deb2013iterative}~\cite{chin2008}, which allows the large ranges of parameter space to be excluded from the overall localization.  While iterative MLE localization is much faster than standard MLE search, throwing out large areas of parameter space makes the search liable to be caught within local maxima~\cite{Sheng2005}.  In~\cite{CordoneExpansion2017}, we proposed the use of Grid Expansion to mitigate these types of errors.  With Grid Expansion, the search area is expanded by a given percent between each iteration, allowing the MLE search to span into areas that would have been thrown out by a standard iterative algorithm.  We found that despite a requiring small increase in computation time, the use of Grid Expansion corrected errors in the situations where the search was getting caught within local maxima, while maintaining good localization performance in cases where the error was not occurring.

\subsection{Linear Regression}
\label{sec:LR}

\begin{figure}[h]
	\centering
	\includegraphics[width=\linewidth]{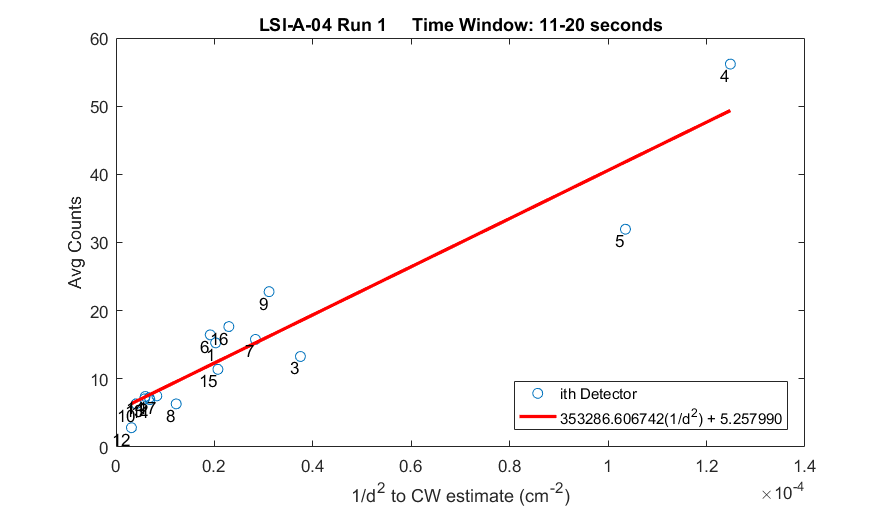}
	\caption{A linear regression model built on average detector measurements vs. their inverse squared distance to an estimate of the source location.  The source and background intensity estimates are provided by the slope and intercept values of the line, which are around $3.533\times10^5$ counts and 5.258 counts, respectively.}
	\label{fig:regressionexample}
\end{figure}

Observe that the radiation propagation model given in (\ref{propagationmodel}) is linear with respect to $1/x^2$.  Furthermore, by the law of large numbers, the average of the Poisson distributed detector measurements over a large time window are representative of the radiation intensity at their respective locations.  Given these two observations, a linear regression model built using the average detector measurements over a large enough time window and their distances to the source location may be used to estimate both the source and background intensities, where the slope of the regression line is the source intensity estimate and the intercept of the regression line is the background intensity estimate.  Figure \ref{fig:regressionexample} shows a linear regression model built using detector data and their inverse-squared distances to a source location estimate over a ten second time window.  In~\cite{CordoneNSS2017}, we use the source estimate from a linear regression model to successfully detect the presence of a moving source within a distributed detector field.

\subsubsection{Linear Regression and MLE}
One of the difficulties of Maximum Likelihood Estimation is the initialization of the search over state space.  In a real time scenario, it is ideal to use small search ranges as a means to conserve computational power and keep the localization up to date with the influx of detector data. While the search area for the position parameters is typically well defined, the search range over the source intensity parameter is not nearly as obvious and can span a large range of possible values.   Furthermore, an MLE localization requires prior knowledge of the background intensity, $B$, since computation of the likelihood function (\ref{loglikelihood}) requires the solution of the propagation model (\ref{propagationmodel}).  In~\cite{CordoneMLERegression}, we show that the source intensity and background estimates provided by the linear regression model can be used to speed up an MLE localization by reducing the search range over the intensity parameter and allowing the removal of detectors from the localization whose measurements are most likely to only include background noise.

\section{Conclusions}
\label{sec:Conc}

In this chapter we highlighted the use of Markov models and stochastic signal processing to learn, extract, fuse, and detect patterns in raw data.  In Section \ref{sec:MM}, we introduced the deterministic Hidden Markov Model (HMM), which is a useful tool to draw information out of a large amount of data.  We described the properties of HMM's and highlighted the usefulness of HMM's for several detection applications. In Section \ref{sec:SS}, we described the uses of stochastic signal processing for the detection and localization of radiation sources.  We highlighted the advantages and drawbacks of localization with Maximum Likelihood Estimation (MLE) and described the estimation of source and background intensities using a linear regression model based on detector counts.

%

\bibliographystyle{spmpsci}
\bibliography{ref}

\end{document}